\documentclass[aps,prd,letterpaper,twocolumn,nofootinbib,preprintnumbers,superscriptaddress,floatfix]{revtex4}

\pdfoutput=1

\usepackage{amsmath,amssymb}
\usepackage{graphicx,color}
\usepackage{multirow}
\usepackage{ulem}

\begin{document}

\preprint{ULB-TH/16-05}
%\title{Resonance Search from Quantum Interference }
\title{Interference Effect on Resonance Studies in Searches of Heavy Particles}
\author{Ligong Bian}
%\email{lgb@itp.ac.cn}
\email{lgbycl@cqu.edu.cn}
\affiliation{Department of Physics, Chongqing University, Chongqing 401331, China}
\affiliation{State Key Laboratory of Theoretical Physics and \\ Kavli Institute for Theoretical Physics China (KITPC), \\ Institute of Theoretical Physics, Chinese Academy of Sciences, Beijing 100190, P. R. China}
\author{Da Liu}
\email{liuda@itp.ac.cn}
\affiliation{State Key Laboratory of Theoretical Physics and \\ Kavli Institute for Theoretical Physics China (KITPC), \\ Institute of Theoretical Physics, Chinese Academy of Sciences, Beijing 100190, P. R. China}
\author{Jing Shu}
\email{jshu@itp.ac.cn}
\affiliation{State Key Laboratory of Theoretical Physics and \\ Kavli Institute for Theoretical Physics China (KITPC), \\ Institute of Theoretical Physics, Chinese Academy of Sciences, Beijing 100190, P. R. China}
\affiliation{CAS Center for Excellence in Particle Physics, Beijing 100049, China}
\author{Yongchao Zhang}
\email{yongchao.zhang@ulb.ac.be}
\affiliation{State Key Laboratory of Theoretical Physics and \\ Kavli Institute for Theoretical Physics China (KITPC), \\ Institute of Theoretical Physics, Chinese Academy of Sciences, Beijing 100190, P. R. China}
\affiliation{Service de Physique Th\'{e}orique, Universit\'{e} Libre de Bruxelles, Boulevard du Triomphe, CP225, 1050 Brussels, Belgium}
\date{\today}

\begin{abstract}
  The interference between resonance signal and continuum background can be either constructive or destructive, depending on the relative sign of couplings between the signal and background amplitudes. Different interference schemes lead to asymmetric distortions of the resonance line shape, which could be distinguished in experiments, when the internal resonance width is  larger than the detector resolution. Interpreting the ATLAS diboson excesses by means of a toy $W'$ model as an illustrative example (though it is disfavored by the 13 TeV data), we find that the signs of resonance couplings can only be revealed in the line shape measurements up to a high confidence level at a high luminosity, which could bring us further information on the underlying theory beyond resonance searches at future lepton and hadron colliders.
  %The ATLAS diboson excesses are viable candidates to see such a line shape difference from reversing the sign of resonance couplings.
  %Interpreting the $WZ$ excess events by means of a toy $W'$ model, we find that the current $\sim$ 2 TeV data show no significant preference for either of the two interference schemes, due to the limited statistics and large reducible non-interfering background. However,  if the diboson excesses are confirmed by the upcoming 14 TeV LHC data,
  %the 3000 fb$^{-1}$ luminosity, even more than $5\sigma$ in the trileptonic $WZ$ decay channel, therefore giving us further information on the underlying theory.
\end{abstract}

\maketitle

%\noindent
\section{Introduction}

The presence of a massive particle always manifests itself at high energy colliders as a resonance peak or significant excess over the (smooth) continuum background, if the decay products from the resonance particle can be reconstructed to some extent. The bound states of heavy quarks, top quark, $W$ and $Z$ bosons, and the 125 GeV standard model (SM) like Higgs are all observed in such a quantum manner. Taking into consideration both the production and decay processes, different helicity states of the resonance particle might interfere with each other~\cite{Buckley:2007th,Buckley2, Keung:2008ve, Murayama:2014yja, Cao:2009ah}. It is more common that the the signal resonance interferes with the continuum background, as on colliders  backgrounds are always unavoidable (non-trivial phase between the signal and background amplitudes could potentially affect dramatically the final observations~\cite{Keung:2008ve, Murayama:2014yja, Jung:2015gta,Choudhury:2011cg,Pappadopulo:2014qza}). Representative examples of such category are the bound on the Higgs total width in the $ZZ^*$ channel \cite{Caola:2013yja, Khachatryan:2014iha, Englert:2014aca, Cacciapaglia:2014rla, Englert:2015zra,Kauer:2015hia,Kauer:2012hd} and the Higgs diphoton channel at the large hadron collider (LHC)~\cite{SM-Higgs:diphoton1,SM-Higgs:diphoton2,SM-Higgs:diphoton3,SM-Higgs:diphoton4,SM-Higgs:diphoton5}, which are two of the primary channels to observe the Higgs particle and precisely determine its mass.

The signal-background interference terms  are subject to the  magnitudes of couplings of the resonance state to the initial and final states (compared to couplings in the background processes) and the resonance width. A wide resonance decay width would generally augment the resonance signal, enlarge the relative size of signal-background interference terms and reduce the detector smearing effect on the resonance line shape. The signs of resonance couplings, more  properly the relative sign between the signal and background amplitudes, do also matter. In case of the same (opposite) sign scenarios, the signal and background amplitudes are additive (subtractive) and interfere with each other constructively (destructively). Combining both the effects from the magnitudes and signs of resonance couplings, the signal-background interference generally leads to distortions of the pure resonance  to some extent. In turn, experimental data in vicinity of the resonance could constrain both the magnitudes and signs of couplings involved and help to discriminate the constructive interference from the destructive one.

With regard to direct searches for heavy states at the current running LHC II and future higher energy colliders, the constructive/destructive interference can be used to examine some specific beyond SM models and  exclude large portions of parameter space. An example at hand is the tantalizing diphoton excesses at 750 GeV~\cite{diphoton}, see e.g.~\cite{Jung:2015etr}. However, the events with photon final states are in general much less than in other channels (or it is very easy to see such high energy photons due to the clean background), and therefore it is rather challenging to extract useful information from line shape measurements even at a realistically large luminosity. As we will see below, only with a huge number of signal events could the resonance shape be used to constrain relevant couplings or beyond SM physics. Thus we resort to the excesses around $\sim$ 2 TeV~\cite{Aad:2015owa} which have a significantly larger cross section than the diphoton events, though it is disfavored by the current 13 TeV data. The analysis in this note is only an illustrative example to reveal how to use resonance-background to constrain new physics; even if the diboson data are falsified by upcoming 13 TeV data, we can still apply such methods to heavy particle searches at future colliders, as long as the requirement of large luminosity can be achieved.

Regarding the resonance at 2 TeV, the most significant hint is in the $WZ$ channel from the ATLAS data~\cite{Aad:2015owa}. These high mass excesses have triggered intensive discussions and interpretations in terms of various beyond SM scenarios~\cite{Bian:2015ota,Thamm2015, diboson1,diboson2,diboson3,diboson4,diboson5,diboson6,diboson7,diboson8,diboson9,diboson10, diboson11,diboson12,diboson13,diboson14,diboson15,diboson16,diboson17,diboson18,diboson19,diboson20, diboson21,diboson22,diboson23,diboson24,diboson25,diboson26,diboson27,diboson28,diboson29,diboson30, diboson31,diboson32,diboson33,diboson34,diboson35,diboson36,diboson37,diboson38,diboson39,diboson40, diboson41,diboson42}. In this work we use the ATLAS $WZ$ excess as trial data to demonstrate the constructive/destructive signal-background interference effect in the framework of a toy $W'$ model, which can be generalized to more realistic scenarios, more intricate analysis and potentially even more resonance-like excesses in the future, in a straightforward way.

\begin{figure}[t]
  %\hspace*{-1.8cm}
  \includegraphics[angle=0,scale=0.5]{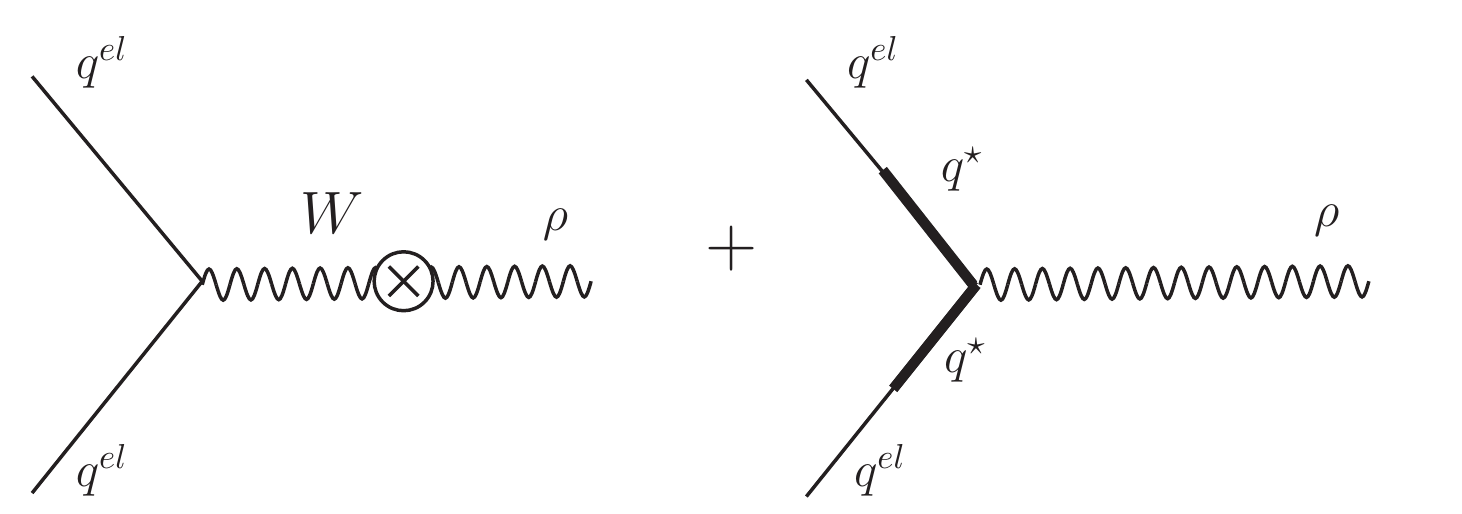}
  \vspace{-0.2cm}
  \caption{
  \label{fig:qrW}
  Two sources of the couplings between $\rho$ and  quarks in the partial compositeness scenario. Left: from the mixing of  $\rho$  with $W$ boson. Right: from the mixing of quarks with their composite partners. }
  \vspace{-0.3cm}
\end{figure}

A realistic example for the diboson excess with both signs of couplings is the $\rho$ boson in composite Higgs models~(see for instance Refs.~\cite{composite,Pappadopulo:2014qza,Bian:2015ota,Thamm2015}). To account for the diboson excess and satisfy the bounds from other channels (mainly the  leptonic channels), the new particle should interact strongly with $WZ$ (mainly the longitudinal  components) and have suppressed couplings to the SM leptons, which can be naturally realized by the  $SU(2)_L$ triplet spin-1 resonance $\rho$ in composite Higgs models. Moreover, by adding some degree of compositeness to the valence quarks (see Fig.~\ref{fig:qrW}), one can tune the couplings of $\rho$ to  quarks and obtain different signs, hence producing the constructive or destructive interference effect.

\section{General Analysis}

In light of completeness of the SM blocks, the presence of new heavy resonance states undoubtedly means the existence of beyond SM new particles and new interactions connecting them to the established fundamental elements. For concreteness, we consider a resonance $X$ decaying into two particles $A$ and $B$, where $A$ and $B$ are any species of (identical) particles within or beyond the SM, and the invariant mass $M_{AB}$ can be (partially) reconstructed at colliders. The amplitude $X \rightarrow AB$ can then be formally cast into the expression
\begin{eqnarray}
\mathcal{M}^S_{X(AB)} = - \frac{ \mathcal{M}^{\rm prod}_{X} \mathcal{M}_{X \rightarrow AB}  }{M_{AB}^2 - M_X^2 + i M_X  \Gamma_X } \,,
\end{eqnarray}
where $M_{X}$ and $\Gamma_X$ are, respectively, the mass and width of $X$, $\mathcal{M}^{\rm prod}_X$ and $\mathcal{M}_{X \rightarrow AB}$ the production and decay amplitudes. The propagator of $X$ has been explicitly shown, which is a crucial factor for the interference phenomena. In vicinity of the resonance, different prescriptions of the resonance structure lead to small discrepancies quantitatively, which becomes more significant when the width goes larger~\cite{Englert:2015zra}. As a viable approximation, we neglect such subtleness and work only in the standard Breit-Wigner formalism throughout this paper. In the mean time, the smooth background $\mathcal{M}_{\rm bkg}$ depends also on the invariant mass $M_{AB}$. In terms of the cross section, the signal goes like
\begin{eqnarray}
\sigma^{S}_{X(AB)}
&=& \int d M_{AB} \frac{S}{(M^2_{AB} - M_X^2)^2 + M_X^2 \Gamma_X^2} \nonumber \\
&\sim& \sigma^{S}_X {\rm BR}( X \rightarrow AB ) \,,
\end{eqnarray}
where $S$ is some factor independent of the resonance propagator, and BR the branching ratio. On the other hand, the integrated interfering cross section reads
\begin{eqnarray}
\sigma^{\rm int}_{X(AB)} = -2  \int d M_{AB} \frac{(M^2_{AB} - M_X^2) \Re + M_X \Gamma_X \Im}{(M^2_{AB} - M_X^2)^2 + M_X^2 \Gamma_X^2} \ ,
\end{eqnarray}
where
\begin{eqnarray}
\Re &\equiv& {\rm Re} (\mathcal{M}^{\rm prod}_{X} \mathcal{M}_{X \rightarrow AB} \mathcal{M}^*_{\rm bkg}) \,, \\
\Im &\equiv& {\rm Im} (\mathcal{M}^{\rm prod}_{X} \mathcal{M}_{X \rightarrow AB} \mathcal{M}^*_{\rm bkg})
\end{eqnarray}
are, respectively, the real and imaginary contributions. We can see that in the on-shell region, the interference terms depend both on $\Re$ and $\Im$ as well as the width $\Gamma_X$. When the invariant mass is far away from the resonance, i.e. $|M^2_{AB} - M_X^2| \gg M_X \Gamma_X$, on the other hand,
only the real part $\Re$ contribute and it goes like
\begin{eqnarray}
\sim -2  \frac{ {\rm Re} (\mathcal{M}^{\rm prod}_{X} \mathcal{M}_{X \rightarrow AB} \mathcal{M}^*_{\rm bkg}) }{ M_{AB}^2 - M_X^2 }  \,.
\end{eqnarray}

In a large variety of popular new physics models, the couplings of resonance $X$ to the decay products and/or the initial particles can take both positive and negative values. Consequently, the signal resonance can interfere with the continuum background constructively or destructively, depending on whether the signal and background amplitudes are additive or subtractive, as aforementioned. Given different signs of the couplings involved, say $\pm g_{XAB}$, the total cross sections $\sigma(pp \rightarrow X \rightarrow AB)$ are generally different, especially when the coupling is small such that the quadratic or higher order terms of $g_{XAB}$ in the cross section are not important~\cite{Bian:2015zha}.

As stated above, the signs of couplings could also change the line shape of $M_{AB}$. Specifically, the constructive interference tends to produce more events in the higher mass region $M_{AB} > M_X$ and, as a result, shift the peak to the upward direction to some extent, while the destructive interference distorts the resonance shape in a right opposite manner. To quantify the asymmetric effect, we define the parameter~\cite{Lillie:2007ve}
\begin{gather}
\label{eq:Ai}
A_{i} \equiv
\frac{\int {\rm d} M_{AB} \left[ \frac{{\rm d} \sigma}{{\rm d} M_{AB}} - \left(\frac{{\rm d} \sigma}{{\rm d} M_{AB}}\right)_{\rm bkg} \right] \Theta(M_{AB} - M_{X} )}
{\int {\rm d} M_{AB} \left| \frac{{\rm d} \sigma}{{\rm d} M_{AB}} - \left( \frac{{\rm d} \sigma}{{\rm d} M_{AB}}\right)_{\rm bkg} \right|} \,,
\end{gather}
where the $\Theta$-function is defined as
\begin{eqnarray}
\Theta(x) \equiv \left\{
\begin{array}{cc}
-1,  & x < 0 \\
  1, & x > 0
\end{array}
\right.
\end{eqnarray}
which changes the sign when the resonance is crossed. The $A_i$ parameter could be either positive or negative depending on the signs of resonance couplings and vanishes for the pure background. The background contribution has been subtracted to determine if the interference is constructive or destructive.\footnote{Note that the parameter $A_i$ is highly nontrivial in the experimental side. The background has to understood very well, and experimentally the mass $M_X$ is not known. In vicinity of the resonance the peak shift due to signal background interference has to been taken into consideration in a proper way. In the analysis below we assume the central bin, c.f. Figure~\ref{fig:inf}, can be well identified and the shift effect is small and can be neglected as a viable approximation.} Note that with this definition, the sign of $A_i$ coincides with the relative sign between the signal and background amplitudes; in other words, $A_i >0 \, (A_i<0)$ indicates the occurrence of constructive (destructive) interference.

Though the magnitudes of the numerator and denominator of $A_i$ depend on the binning of data in vicinity of the resonance, the sign of $A_i$ does not. One can extract the $A_i$ parameter from experimental data, and also obtain it from some underlying theories or models with different signs of couplings, say $\pm g_{XAB}$ for the $X$ resonance. By comparing the values of $A_i$ from experimental data and theoretical predictions, one can infer in a straightforward way which interference scheme is preferred, thus constraining the couplings and parameter space in some specific models.

%\section{2 TeV diboson excess }
\section{Limited statistics}
\label{sec:diboson}

We utilize a toy $W^\prime$ model to test the constructive/destructive signal-background interference in the ATLAS $WZ$ channel. It is expected that with the limited statistics and large background we can not have any significant hints of the signs of couplings.

Assuming simply the generation-universal coupling $g_{W' ud}$ and the $W'WZ$ coupling coefficient $g_{W'WZ}$ in terms of the SM $WWZ$ interaction, the toy $W'$ model can in some sense mimic the extra charged gauge boson in left-right symmetric models~\cite{LRSM1,LRSM2,LRSM3} or the $\rho^\pm$ boson in composite Higgs models. To be more specific, we fix the mass $M_{W'} = 2$ TeV, the width $\Gamma_{W'} = 70$ GeV and $g_{W'WZ} = +0.005$~\cite{Bian:2015ota,Thamm2015}, then the constructive and destructive interference scenarios emerge as the coupling $g_{W^\prime ud}$ being, respectively, $\pm 0.15$. The signal process $pp \rightarrow W' \rightarrow WZ$ interferes with the SM background $pp \rightarrow WZ$, with the $W'$ boson naturally the origin of $\sim$ 2 TeV $WZ$ excess events.

We implement simple cuts on the fat $W$ and $Z$ jets: $p_T > 540$ GeV and $|\eta| < 2$. The smearing effect due to the finite detector resolution of the momenta of jets is  taken into consideration, following the procedure in~\cite{Aad:2015owa}. Following Ref.~\cite{ATLAS:combined}, we assume the signal acceptance times efficiency factor of $\epsilon \simeq 0.07$. Given the benchmark values of input parameters for the $W'$ models given above, we can fit roughly the ATLAS $WZ$ excess~\cite{Pukhov:2004ca}. Due to the limited statistics of current data, we use only the three bins from 1.85 to 2.15 TeV to calculate $A_i$, both for the constructive/destructive interference scenarios and the ATLAS $WZ$ data, which come out to be $\pm 0.11$ and $-0.52^{+1.52}_{-0.48}$. From these values we can see that the destructive interference is relatively preferred by the central value of $A_i$ from current data. However, as a result of the low statistics and large reducible non-interfering $JJ$ background, we can not distinguish clearly the two interference schemes. In the fit, we find that the jet smearing effect can moderately broaden the  resonance and tend to slightly decrease the difference of $A_i$ for resonance couplings with different signs.

%\begin{table}[tp]
%  \caption{
%  \label{tab:intf}
%  Input parameters for the constructive and destructive interference schemes of the toy $W'$ model.}
%  \begin{tabular}{lcccc}
%  \hline\hline
%  interference & Mass  & $\Gamma_{W'}$  &  $g_{W^\prime WZ}$& $g_{W^\prime ud}$  \\
%  \hline
%  constructive & 2 TeV   & 70 GeV & 0.005 & $+0.15$  \\
%  destructive  & 2 TeV   & 70 GeV & 0.005 & $-0.15$  \\
%  \hline\hline
%  \end{tabular}
%\end{table}

%\begin{table}[t]
%  \caption{
%  \label{tab:asym}
%  Local $A_i$ for the constructive/destructive interference schemes with input parameters given in Table~\ref{tab:intf} and the 8 TeV ATLAS $WZ$ data. }
%  \begin{tabular}{lccc}
%  \hline\hline
%     & constructive     & destructive & data  \\
%  \hline
%     $A_i$ & 0.11  & $-0.11$ & $-0.52^{+1.52}_{-0.48}$ \\
%  \hline\hline
%  \end{tabular}
%  \vspace{-0.3cm}
%\end{table}

%\section{Prospects at LHC run II}
\section{Prospects at large luminosity}

It is promising that the constructive and destructive interference hypotheses can be more clearly differentiated at the higher energy, say LHC run II, with much more signal data. It is promising that with upcoming more data at 13 TeV LHC, we can have soon decisive conclusion on the resonance at 2 TeV. As an explicit example, we examine the signal background interference effect for the 2 TeV resonance at 14 TeV with a large luminosity.

%To examine the prospects for observing the $\sim$ 2 TeV resonance, we estimate roughly the numbers of signal and dominant background events at 14 TeV LHC with an integrated luminosity of 20 fb$^{-1}$, which are collected in Table~\ref{tab:prospects}.

%and count only the events in the local bin $[1950, 2050]$ GeV.
%In absence of the counterpart data at 13 or 14 TeV,

To be concrete, we utilize the input parameters in the last section. All the four channels of leptonic, semileptonic, and hadronic decays are considered: $\ell\ell\ell'\nu$, $\ell\ell q\bar{q}'$, $\ell\nu q\bar{q}$ and $q\bar{q}q'\bar{q}^{\prime\prime}$ with $\ell,\, \ell' = e,\, \mu$. As the decay products are always highly boosted, it is common that some of the $q\bar{q}^{(\prime)}$ events appear to be large-$R$ jets. We simulate the signal process $pp \rightarrow W' \rightarrow WZ$ and the dominate backgrounds in Table~\ref{tab:prospects}, and rescale simply the current 8 (or 13) TeV data to 14 TeV, assuming na\"ively the event efficiencies being the same at the two energy scales. The signal acceptance times efficiency for the four distinct channels are from Fig.~1(a) of~\cite{ATLAS:combined}, where the branching fractions of $W/Z$ decays have been taken into consideration. The background simulations follow Refs~\cite{atlas1,atlas2,atlas3,Aad:2015owa,Khachatryan:2014gha}, for which we implement only the basic event selection cuts. It should be aware that in the high mass region all these channels might suffer from large systematic and/or statistical uncertainties, depending on the future high energy data. In simulations we find that the hadronic channel is the most promising to confirm or exclude the 2 TeV resonance, due to the large branching ratio.
%The expected local significances for the four channels are listed in the last column of Table~\ref{tab:prospects}, where the standard Poisson distribution is implemented. It is obvious that the hadronic channel can see most events and might be most promising to confirm the 2 TeV resonance.

%\begin{table}[t]
%  \caption{
%  \label{tab:prospects}
%  Prospects for the 2 TeV resonance in the $WZ$ channel at the 14 TeV LHC, assuming a luminosity of 20 fb$^{-1}$. The second column lists the dominant reducible/irreducible backgrounds, cf. Refs~\cite{ATLAS:combined,atlas1,atlas2,atlas3,Khachatryan:2014gha}. The following columns list the expected numbers of signals and backgrounds and the corresponding local significance in the invariant mass window of [1.95, 2.05] TeV.  }
%  \begin{tabular}{ccccc}
%  \hline\hline
%     channel & backgrounds & $S$ & $B$ & significance  \\
%  \hline
%  $\ell\ell\ell'\nu$ & $WZ$ & 6.6 & 1.9 & 3.2  \\
%  $\ell\ell q\bar{q}'$ & $Z+$jets & 13.1  & 4.9 & 4.5  \\
%  $\ell\nu q\bar{q}$  & $W/Z+$jets & 39.4 & 14.7 & 7.8  \\
%  $q\bar{q}q'\bar{q}^{\prime\prime}$ & $jj$ & 46.0  & 8.8 & 10.4  \\
%  \hline\hline
%  \end{tabular}
%  \vspace{-0.3cm}
%\end{table}

\begin{table}[t]
  \caption{
  \label{tab:prospects}
  Signals and dominate reducible/irreducible backgrounds in the $WZ$ final states at the 14 TeV LHC~\cite{ATLAS:combined,atlas1,atlas2,atlas3,Khachatryan:2014gha}.  }
  \begin{tabular}{cc}
  \hline\hline
     channel & backgrounds  \\
  \hline
  $\ell\ell\ell'\nu$ & $WZ$  \\
  $\ell\ell q\bar{q}'$ & $Z+$jets \\
  $\ell\nu q\bar{q}$  & $W/Z+$jets   \\
  $q\bar{q}q'\bar{q}^{\prime\prime}$ & $jj$ \\
  \hline\hline
  \end{tabular}
  \vspace{-0.3cm}
\end{table}

\begin{figure}[t]
  %\hspace*{-1.8cm}
  \includegraphics[angle=0,scale=0.6]{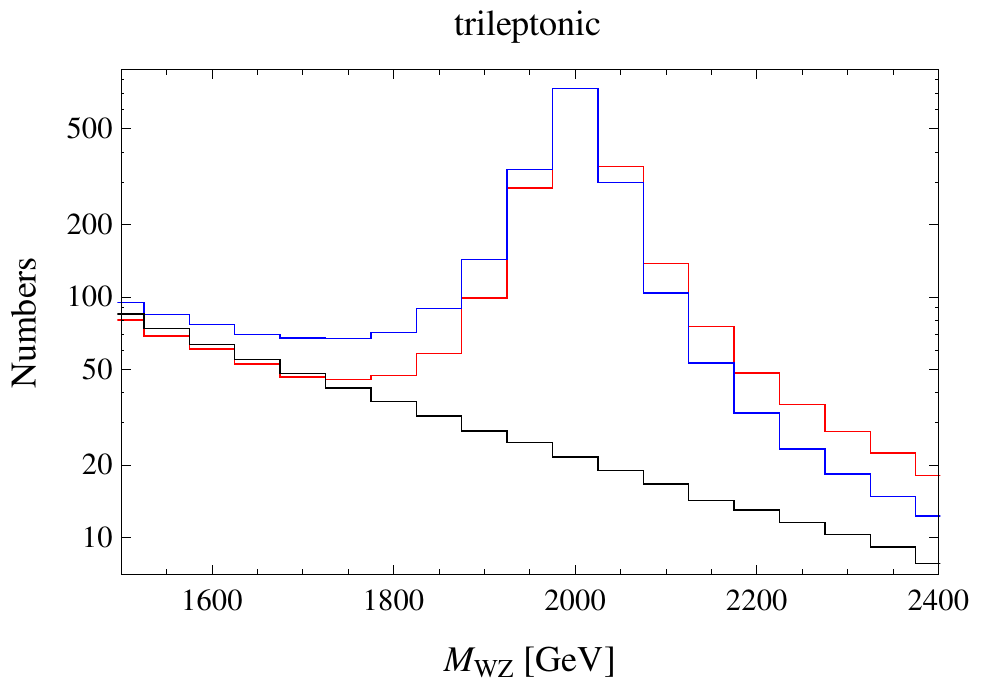} \\
  \includegraphics[angle=0,scale=0.6]{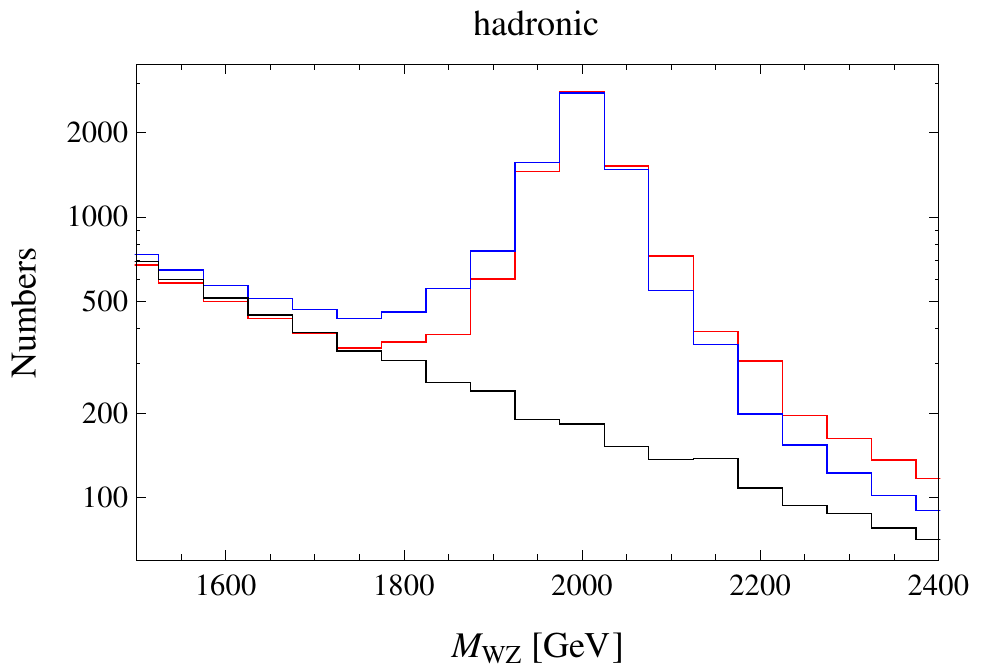}
  \caption{\label{fig:inf}
  Invariant mass distribution of $M_{WZ}$ at the 14 TeV LHC: the dark, orange and blue lines indicate, respectively, the simulated background and resonances with constructive and destructive interference at the integrated luminosity of 3000 fb$^{-1}$.}
  \vspace{-0.3cm}
\end{figure}

As an explicit example of the interference effect, we show in Fig.~\ref{fig:inf} the invariant mass $M_{WZ}$ in the trileptonic and hadronic channels at the 14 TeV LHC. The simulated line shapes for the background and the constructive/destructive interfering resonances are shown, respectively, as dark, orange/blue lines, assuming a total luminosity of 3000 fb$^{-1}$. We reduce the rescaled $JJ$ background in the hadronic channel from the 8 TeV data by a factor of two, assuming for more events at LHC Run II more jet techniques are used and a more aggressive cut is made. For simplicity we implement only the simple cuts as did in~\cite{atlas1} and~\cite{Aad:2015owa}. The quark jet smearing is performed as stated above, while for the charged leptons we assume na\"ively the energy uncertainty $\Delta E / E \simeq 1\%$~\cite{Aad:2009wy}. It is found that the lepton smearing have only tiny effect on the $M_{WZ}$ line shape and $A_i$.
%The input parameters for the leptonic and hadronic channels are the same as given above, and
The expected local $A_i$ for the two channels are presented in Table~\ref{tab:asym2}.\footnote{Notice that large $JJ$ background in the hadronic channel does not contribute to the $A_i$ factor, however, in real data analysis, the effect of backgrounds on the uncertainties of $A_i$ has to be taken into consideration.}

For the ``standard'' scenarios Table~\ref{tab:asym2}, the central 17 bins (with a bin width of 50 GeV) around 2 TeV are used to calculate $A_i$. For the ``optimal" scenarios, on the other hand, the central seven bins are removed in the calculation from the 17 bins, so only the $5+5=10$ bins with larger interference effect are used and we obtain the more aggressive predictions for the asymmetry factor while suffer from larger statistical uncertainties due to the reduced statistics. The uncertainties in the last column of Table~\ref{tab:asym2} are the purely statistical one by simply counting the event numbers. It is expected that the constructive and destructive interference schemes can be differentiated at a reasonably large confidence level in both channels. In the trileptonic channel it could even be further improved to more than $5\sigma$ in the ``optimal'' scenario. Here we perform only rather na\"ive examinations of the prospects in the purely leptonic and hadronic channels, in the semileptonic channels we also expect a significant differentiation of the two interference schemes. When all these channels are combined together, the significance can even be further improved. Furthermore, more accurate estimations call for much more intricate simulations and analysis. Short in all, in light of the estimated $A_i$ and uncertainties in Table~\ref{tab:asym2}, given a {\it huge} amount of data at the current running LHC II and more refined experimental analysis, for instance the Boosted Decision Trees method~\cite{Roe:2004na}, we could probably reduce the statistical and systematic errors to a sufficiently low level such that we can measure the asymmetry factor $A_i$ precisely and pin down which interference hypothesis is the truth for the 2 TeV resonance, and thus constrain the signs of beyond SM couplings.

\begin{table}[t]
  \caption{
  \label{tab:asym2}
  Expected local $A_i$ in the trileptonic and hadronic channels for the constructive/destructive interference schemes at the 14 TeV LHC, and their corresponding statistical uncertainties at the integrated luminosity of 3000 fb$^{-1}$. }
  \begin{tabular}{llccc}
  \hline\hline
     $A_i$  & scenario & constructive     & destructive & uncertainty  \\
  \hline
    \multirow{2}{*}{trileptonic$\;$} & standard   & 0.25  & $-0.13$ & 0.09 \\ \cline{2-5}
                                 & optimal   & 0.77  & $-0.37$ & 0.18 \\  \hline
    \multirow{2}{*}{hadronic}    & standard  & 0.20  & $-0.10$ & 0.12  \\ \cline{2-5}
                                 & optimal  & 0.79  & $-0.33$ & 0.54  \\
  \hline\hline
  \end{tabular}
  \vspace{-0.3cm}
\end{table}

\section{Conclusions}
\label{sec:conclusions}

Quantum interference is very common in the regime of elementary particle physics. In presence of some resonances on top of the continuum background, it is unavoidable that interference would occur between the signal and background. The shape of resonances depends both on the magnitudes of couplings involved and on the relative sign between the signal and background terms, i.e. whether the interference is constructive or destructive. In this work we point out how the resonance shape is affected and how to use the asymmetry parameter $A_i$ to differentiate the two distinct interference schemes.

Though a 2 TeV resonance is not favored by current 13 TeV data, the ATLAS diboson excesses are a viable illustrative candidate for the time being to test the implications of interference phenomena for future searches and studies of high mass resonances at LHC run II and the next-generation higher energy colliders. Implementing a toy $W'$ model as a solution to the excess in $WZ$ channel, we find that the constructive and destructive interference schemes could be differentiated at a reasonably large confidence level in the trileptonic and hadronic $WZ$ decay channels (and also possible in the semileptonic channels), and further improved to more than $5\sigma$ in the ``optimal'' scenario for the trileptonic channel, as long as a huge statistics of signal events is achieved (a luminosity of 3000 fb$^{-1}$ assumed for the diboson events). Even if the 2 TeV excesses are excluded by upcoming data, the signal-background interference and resonance line shape are always useful to constrain the magnitudes and signs of beyond SM couplings and exclude large portions of parameter space in specific models.

\section*{Acknowledgements}
We would like to thank Qing-Hong Cao, Lian-Tao Wang for the illustrating discussions. YZ would also like to thank Jean-Marie Fr\`{e}re for enlightening comments on the manuscript and acknowledgement the valuable discussions with Roman Kogler.  This work is supported by the Fundamental Research Funds for the Central Universities (under grant No. 0903005203404). The work of YZ is partially supported by the IISN and Belgian Science Policy (IAP VII/37).


\begin{thebibliography}{99}

\bibitem{Buckley:2007th}
  M.~R.~Buckley, H.~Murayama, W.~Klemm and V.~Rentala,
  %``Discriminating spin through quantum interference,''
  Phys.\ Rev.\ D {\bf 78}, 014028 (2008)
  [arXiv:0711.0364 [hep-ph]].
\bibitem{Buckley2}
  M.~R.~Buckley, B.~Heinemann, W.~Klemm and H.~Murayama,
  %``Quantum Interference Effects Among Helicities at LEP-II and Tevatron,''
  Phys.\ Rev.\ D {\bf 77}, 113017 (2008)
  [arXiv:0804.0476 [hep-ph]].

%\cite{Keung:2008ve}
\bibitem{Keung:2008ve}
  W.~Y.~Keung, I.~Low and J.~Shu,
  %``Landau-Yang Theorem and Decays of a Z Boson into Two Z Bosons,''
  Phys.\ Rev.\ Lett.\  {\bf 101}, 091802 (2008)
  [arXiv:0806.2864 [hep-ph]].
%\cite{Cao:2009ah}
\bibitem{Cao:2009ah}
  Q.~H.~Cao, C.~B.~Jackson, W.~Y.~Keung, I.~Low and J.~Shu,
  %``The Higgs Mechanism and Loop-induced Decays of a Scalar into Two Z Bosons,''
  Phys.\ Rev.\ D {\bf 81}, 015010 (2010)
  [arXiv:0911.3398 [hep-ph]].
  %%CITATION = ARXIV:0911.3398;%%
  %46 citations counted in INSPIRE as of 08 sept. 2015
%\cite{Murayama:2014yja}
\bibitem{Murayama:2014yja}
  H.~Murayama, V.~Rentala and J.~Shu,
  %``Disentangling Strong Dynamics through Quantum Interferometry,''
  arXiv:1401.3761 [hep-ph].
  %%CITATION = ARXIV:1401.3761;%%
  %3 citations counted in INSPIRE as of 08 sept. 2015

\bibitem{Jung:2015gta}
  S.~Jung, J.~Song and Y.~W.~Yoon,
  %``Dip or nothingness of a Higgs resonance from the interference with a complex phase,''
  arXiv:1505.00291 [hep-ph].

\bibitem{Choudhury:2011cg}
  D.~Choudhury, R.~M.~Godbole and P.~Saha,
  %``Dijet resonances, widths and all that,''
  JHEP {\bf 1201}, 155 (2012)
  [arXiv:1111.1054 [hep-ph]].

\bibitem{Pappadopulo:2014qza}
  D.~Pappadopulo, A.~Thamm, R.~Torre and A.~Wulzer,
  %``Heavy Vector Triplets: Bridging Theory and Data,''
  JHEP {\bf 1409}, 060 (2014)
  [arXiv:1402.4431 [hep-ph]].

%\cite{Caola:2013yja}
\bibitem{Caola:2013yja}
  F.~Caola and K.~Melnikov,
  %``Constraining the Higgs boson width with ZZ production at the LHC,''
  Phys.\ Rev.\ D {\bf 88}, 054024 (2013)
  [arXiv:1307.4935 [hep-ph]].
  %%CITATION = ARXIV:1307.4935;%%
  %95 citations counted in INSPIRE as of 08 sept. 2015

%\cite{Khachatryan:2014iha}
\bibitem{Khachatryan:2014iha}
  V.~Khachatryan {\it et al.} [CMS Collaboration],
  %``Constraints on the Higgs boson width from off-shell production and decay to Z-boson pairs,''
  Phys.\ Lett.\ B {\bf 736}, 64 (2014)
  [arXiv:1405.3455 [hep-ex]].
  %%CITATION = ARXIV:1405.3455;%%
  %89 citations counted in INSPIRE as of 08 sept. 2015

%\cite{Englert:2014aca}
\bibitem{Englert:2014aca}
  C.~Englert and M.~Spannowsky,
  %``Limitations and Opportunities of Off-Shell Coupling Measurements,''
  Phys.\ Rev.\ D {\bf 90}, 053003 (2014)
  [arXiv:1405.0285 [hep-ph]].
  %%CITATION = ARXIV:1405.0285;%%
  %43 citations counted in INSPIRE as of 08 sept. 2015
%\cite{Cacciapaglia:2014rla}
\bibitem{Cacciapaglia:2014rla}
  G.~Cacciapaglia, A.~Deandrea, G.~Drieu La Rochelle and J.~B.~Flament,
  %``Higgs couplings: disentangling New Physics with off-shell measurements,''
  Phys.\ Rev.\ Lett.\  {\bf 113}, no. 20, 201802 (2014)
  [arXiv:1406.1757 [hep-ph]].
  %%CITATION = ARXIV:1406.1757;%%
  %19 citations counted in INSPIRE as of 08 sept. 2015

\bibitem{Englert:2015zra}
  C.~Englert, I.~Low and M.~Spannowsky,
  %``On-shell interference effects in Higgs boson final states,''
  Phys.\ Rev.\ D {\bf 91}, no. 7, 074029 (2015)
  [arXiv:1502.04678 [hep-ph]].

\bibitem{Kauer:2015hia}
  N.~Kauer and C.~O¡¯Brien,
  %``Heavy Higgs signal¨Cbackground interference in $gg\rightarrow VV$ in the Standard Model plus real singlet,''
  Eur.\ Phys.\ J.\ C {\bf 75}, 374 (2015)
  [arXiv:1502.04113 [hep-ph]].

\bibitem{Kauer:2012hd}
  N.~Kauer and G.~Passarino,
  %``Inadequacy of zero-width approximation for a light Higgs boson signal,''
  JHEP {\bf 1208}, 116 (2012)
  [arXiv:1206.4803 [hep-ph]].

\bibitem{SM-Higgs:diphoton1}
  D.~A.~Dicus and S.~S.~D.~Willenbrock,
  %``Photon Pair Production and the Intermediate Mass Higgs Boson,''
  Phys.\ Rev.\ D {\bf 37}, 1801 (1988).
\bibitem{SM-Higgs:diphoton2}
  L.~J.~Dixon and M.~S.~Siu,
  %``Resonance continuum interference in the diphoton Higgs signal at the LHC,''
  Phys.\ Rev.\ Lett.\  {\bf 90}, 252001 (2003)
  [hep-ph/0302233].
\bibitem{SM-Higgs:diphoton3}
  S.~P.~Martin,
  %``Shift in the LHC Higgs diphoton mass peak from interference with background,''
  Phys.\ Rev.\ D {\bf 86}, 073016 (2012)
  [arXiv:1208.1533 [hep-ph]].
\bibitem{SM-Higgs:diphoton4}
  L.~J.~Dixon and Y.~Li,
  %``Bounding the Higgs Boson Width Through Interferometry,''
  Phys.\ Rev.\ Lett.\  {\bf 111}, 111802 (2013)
  [arXiv:1305.3854 [hep-ph]].
\bibitem{SM-Higgs:diphoton5}
  G.~Z.~Xu, G.~Li, Y.~J.~Li, K.~Y.~Liu and Y.~J.~Zhang,
  %``Interference effects on Higgs mass measurement in $e^+e^-\to H(\gamma\gamma) Z$ at CEPC,''
  arXiv:1505.06981 [hep-ph].

\bibitem{diphoton}
  The ATLAS collaboration,
  %``Search for resonances decaying to photon pairs in 3.2 fb$^{-1}$ of $pp$ collisions at $\sqrt{s}$ = 13 TeV with the ATLAS detector,''
  ATLAS-CONF-2015-081;
%\bibitem{diphoton2}
  CMS Collaboration [CMS Collaboration],
  %``Search for new physics in high mass diphoton events in proton-proton
  collisions at 13TeV,''
  CMS-PAS-EXO-15-004.

\bibitem{Jung:2015etr}
  S.~Jung, J.~Song and Y.~W.~Yoon,
  %``How Resonance-Continuum Interference Changes 750 GeV Diphoton Excess: Signal Enhancement and Peak Shift,''
  arXiv:1601.00006 [hep-ph].

\bibitem{Aad:2015owa}
  G.~Aad {\it et al.}  [ATLAS Collaboration],
  %``Search for high-mass diboson resonances with boson-tagged jets in proton-proton collisions at $\sqrt{s}$ = 8 TeV with the ATLAS detector,''
  arXiv:1506.00962 [hep-ex].

%\cite{Bian:2015ota}
\bibitem{Bian:2015ota}
  L.~Bian, D.~Liu and J.~Shu,
  %``Low Scale Composite Higgs Model and 1.8 $\sim$ 2 TeV Diboson Excess,''
  arXiv:1507.06018 [hep-ph].
  %%CITATION = ARXIV:1507.06018;%%
  %9 citations counted in INSPIRE as of 08 sept. 2015
\bibitem{Thamm2015}
  A.~Thamm, R.~Torre and A.~Wulzer,
  %``A composite Heavy Vector Triplet in the ATLAS di-boson excess,''
  arXiv:1506.08688 [hep-ph].

\bibitem{diboson1}
  H.~S.~Fukano, M.~Kurachi, S.~Matsuzaki, K.~Terashi and K.~Yamawaki,
  %``2 TeV Walking Technirho at LHC?,''
  arXiv:1506.03751 [hep-ph].
\bibitem{diboson2}
  J.~Hisano, N.~Nagata and Y.~Omura,
  %``Interpretations of the ATLAS Diboson Resonances,''
  arXiv:1506.03931 [hep-ph].
\bibitem{diboson3}
  D.~B.~Franzosi, M.~T.~Frandsen and F.~Sannino,
  %``Diboson Signals via Fermi Scale Spin-One States,''
  arXiv:1506.04392 [hep-ph].
\bibitem{diboson4}
  K.~Cheung, W.~Y.~Keung, P.~Y.~Tseng and T.~C.~Yuan,
  %``Interpretations of the ATLAS Diboson Anomaly,''
  arXiv:1506.06064 [hep-ph].
\bibitem{diboson5}
  B.~A.~Dobrescu and Z.~Liu,
  %``A W' Boson near 2 TeV: Predictions for Run 2 of the LHC,''
  arXiv:1506.06736 [hep-ph].
\bibitem{diboson6}
  J.~A.~Aguilar-Saavedra,
  %``Triboson interpretations of the ATLAS diboson excess,''
  arXiv:1506.06739 [hep-ph].
\bibitem{diboson7}
  A.~Alves, A.~Berlin, S.~Profumo and F.~S.~Queiroz,
  %``Dirac-Fermionic Dark Matter in $U(1)_X$ Models,''
  arXiv:1506.06767 [hep-ph].
\bibitem{diboson8}
  Y.~Gao, T.~Ghosh, K.~Sinha and J.~H.~Yu,
  %``G221 Interpretations of the Diboson and Wh Excesses,''
  arXiv:1506.07511 [hep-ph].
\bibitem{diboson9}
  J.~Brehmer, J.~Hewett, J.~Kopp, T.~Rizzo and J.~Tattersall,
  %``Symmetry Restored in Dibosons at the LHC?,''
  arXiv:1507.00013 [hep-ph].
\bibitem{diboson10}
  Q.~H.~Cao, B.~Yan and D.~M.~Zhang,
  %``Simple Non-Abelian Extensions and Diboson Excesses at the LHC,''
  arXiv:1507.00268 [hep-ph].

\bibitem{diboson11}
  G.~Cacciapaglia and M.~T.~Frandsen,
  %``Unitarity implications of diboson resonance in the TeV region for Higgs physics,''
  arXiv:1507.00900 [hep-ph].
\bibitem{diboson12}
  T.~Abe, R.~Nagai, S.~Okawa and M.~Tanabashi,
  %``Unitarity sum rules, three site moose model, and the ATLAS 2 TeV diboson anomalies,''
  arXiv:1507.01185 [hep-ph].
\bibitem{diboson13}
  J.~Heeck and S.~Patra,
  %``Minimal Left-Right Dark Matter,''
  arXiv:1507.01584 [hep-ph].
\bibitem{diboson14}
  B.~C.~Allanach, B.~Gripaios and D.~Sutherland,
  %``Anatomy of the ATLAS diboson anomaly,''
  arXiv:1507.01638 [hep-ph].
\bibitem{diboson15}
  T.~Abe, T.~Kitahara and M.~M.~Nojiri,
  %``Prospects for Spin-1 Resonance Search at 13 TeV LHC and the ATLAS Diboson Excess,''
  arXiv:1507.01681 [hep-ph].
\bibitem{diboson16}
  A.~Carmona, A.~Delgado, M.~Quiros and J.~Santiago,
  %``Diboson resonant production in non-custodial composite Higgs models,''
  arXiv:1507.01914 [hep-ph].
\bibitem{diboson17}
  B.~A.~Dobrescu and Z.~Liu,
  %``Heavy Higgs bosons and the 2 TeV $W'$ boson,''
  arXiv:1507.01923 [hep-ph].
\bibitem{diboson18}
  C.~W.~Chiang, H.~Fukuda, K.~Harigaya, M.~Ibe and T.~T.~Yanagida,
  %``Diboson Resonance as a Portal to Hidden Strong Dynamics,''
  arXiv:1507.02483 [hep-ph].
\bibitem{diboson19}
  G.~Cacciapaglia, A.~Deandrea and M.~Hashimoto,
  %``A scalar hint from the diboson excess?,''
  arXiv:1507.03098 [hep-ph].
\bibitem{diboson20}
  H.~S.~Fukano, S.~Matsuzaki and K.~Yamawaki,
  %``Conformal Barrier for New Vector Bosons Decay to the Higgs,''
  arXiv:1507.03428 [hep-ph].

\bibitem{diboson21}
  V.~Sanz,
  %``On the compatibility of the diboson excess with a gg-initiated composite sector,''
  arXiv:1507.03553 [hep-ph].
\bibitem{diboson22}
  C.~H.~Chen and T.~Nomura,
  %``2 TeV Higgs boson and diboson excess at the LHC,''
  arXiv:1507.04431 [hep-ph].
\bibitem{diboson23}
  M.~E.~Krauss and W.~Porod,
  %``Is the CMS eejj excess a hint for light supersymmetry?,''
  arXiv:1507.04349 [hep-ph].
\bibitem{diboson24}
  Y.~Omura, K.~Tobe and K.~Tsumura,
  %``Survey of Higgs interpretations of the diboson excesses,''
  arXiv:1507.05028 [hep-ph].
\bibitem{diboson25}
  W.~Chao,
  %``ATLAS Diboson Excesses from the Stealth Doublet Model,''
  arXiv:1507.05310 [hep-ph].
\bibitem{diboson26}
  L.~A.~Anchordoqui, I.~Antoniadis, H.~Goldberg, X.~Huang, D.~Lust and T.~R.~Taylor,
  %``Stringy origin of diboson and dijet excesses at the LHC,''
  arXiv:1507.05299 [hep-ph].
\bibitem{diboson27}
  D.~Kim, K.~Kong, H.~M.~Lee and S.~C.~Park,
  %``ATLAS Diboson Excesses Demystified in Effective Field Theory Approach,''
  arXiv:1507.06312 [hep-ph].
\bibitem{diboson28}
  H.~Fritzsch,
  %``Composite Weak Bosons at the LHC,''
  arXiv:1507.06499 [hep-ph].
\bibitem{diboson29}
  K.~Lane and L.~Prichett,
  %``Heavy Vector Partners of the Light Composite Higgs,''
  arXiv:1507.07102 [hep-ph].
\bibitem{diboson30}
  A.~E.~Faraggi and M.~Guzzi,
  %``Extra $Z^\prime$s and $W^\prime$s in Heterotic--String Derived Models,''
  arXiv:1507.07406 [hep-ph].

\bibitem{diboson31}
  M.~Low, A.~Tesi and L.~T.~Wang,
  %``Composite spin-1 resonances at the LHC,''
  arXiv:1507.07557 [hep-ph].
\bibitem{diboson32}
  S.~P.~Liew and S.~Shirai,
  %``Testing ATLAS Diboson Excess with Dark Matter Searches at LHC,''
  arXiv:1507.08273 [hep-ph].
\bibitem{diboson33}
  H.~Terazawa and M.~Yasue,
  %``Excited Gauge and Higgs Bosons in the Unified Composite Model,''
  arXiv:1508.00172 [hep-ph].
\bibitem{diboson34}
  P.~Arnan, D.~Espriu and F.~Mescia,
  %``Interpreting a 2 TeV resonance in WW scattering,''
  arXiv:1508.00174 [hep-ph].
\bibitem{diboson35}
  C.~Niehoff, P.~Stangl and D.~M.~Straub,
  %``Direct and indirect signals of natural composite Higgs models,''
  arXiv:1508.00569 [hep-ph].
\bibitem{diboson36}
  P.~S.~B.~Dev and R.~N.~Mohapatra,
  %``Unified explanation of the $eejj$, diboson and dijet resonances at the LHC,''
  arXiv:1508.02277 [hep-ph].
\bibitem{diboson37}
  A.~Dobado, F.~K.~Guo and F.~J.~Llanes-Estrada,
  %``Production cross section estimates for strongly-interacting Electroweak Symmetry Breaking Sector resonances at particle colliders,''
  arXiv:1508.03544 [hep-ph].
\bibitem{diboson38}
  P.~Coloma, B.~A.~Dobrescu and J.~Lopez-Pavon,
  %``Right-handed neutrinos and the 2 TeV $W'$ boson,''
  arXiv:1508.04129 [hep-ph].
\bibitem{diboson39}
  S.~Fichet and G.~von Gersdorff,
  %``Effective theory for neutral resonances and a statistical dissection of the ATLAS diboson excess,''
  arXiv:1508.04814 [hep-ph].
\bibitem{diboson40}
  C.~Petersson and R.~Torre,
  %``ATLAS diboson excess from low scale supersymmetry breaking,''
  arXiv:1508.05632 [hep-ph].
\bibitem{diboson41}
  F.~F.~Deppisch, L.~Graf, S.~Kulkarni, S.~Patra, W.~Rodejohann, N.~Sahu and U.~Sarkar,
  %``Reconciling the 2 TeV Excesses at the LHC in a Linear Seesaw Left-Right Model,''
  arXiv:1508.05940 [hep-ph].
\bibitem{diboson42}
  D.~Gon\c{c}alves, F.~Krauss and M.~Spannowsky,
  %``Augmenting the diboson excess for the LHC Run II,''
  Phys.\ Rev.\ D {\bf 92}, no. 5, 053010 (2015)
  [arXiv:1508.04162 [hep-ph]].

\bibitem{composite}
  D.~Marzocca, M.~Serone and J.~Shu,
  %``General Composite Higgs Models,''
  JHEP {\bf 1208}, 013 (2012)
  [arXiv:1205.0770 [hep-ph]].

\bibitem{Bian:2015zha}
  L.~Bian, J.~Shu and Y.~Zhang,
  %``Prospects for Triple Gauge Coupling Measurements at the 14 TeV LHC and Future Lepton Colliders,''
  arXiv:1507.02238 [hep-ph].

\bibitem{Lillie:2007ve}
  B.~Lillie, J.~Shu and T.~M.~P.~Tait,
  %``Kaluza-Klein Gluons as a Diagnostic of Warped Models,''
  Phys.\ Rev.\ D {\bf 76}, 115016 (2007)
  [arXiv:0706.3960 [hep-ph]].

\bibitem{LRSM1}
  J.~C.~Pati and A.~Salam,
  %``Lepton Number as the Fourth Color,''
  Phys.\ Rev.\ D {\bf 10}, 275 (1974)
  [Phys.\ Rev.\ D {\bf 11}, 703 (1975)].
\bibitem{LRSM2}
  R.~N.~Mohapatra and J.~C.~Pati,
  %``Left-Right Gauge Symmetry and an Isoconjugate Model of CP Violation,''
  Phys.\ Rev.\ D {\bf 11}, 566 (1975).
\bibitem{LRSM3}
  G.~Senjanovic and R.~N.~Mohapatra,
  %``Exact Left-Right Symmetry and Spontaneous Violation of Parity,''
  Phys.\ Rev.\ D {\bf 12}, 1502 (1975).

\bibitem{ATLAS:combined}
  ATLAS Collaboration, ATLAS-CONF-2015-045.

%\cite{Pukhov:2004ca}
\bibitem{Pukhov:2004ca}
  A.~Pukhov,
  %``CalcHEP 2.3: MSSM, structure functions, event generation, batchs, and generation of matrix elements for other packages,''
  hep-ph/0412191.
  %%CITATION = HEP-PH/0412191;%%
  %501 citations counted in INSPIRE as of 29 juil. 2015

\bibitem{atlas1}
  G.~Aad {\it et al.} [ATLAS Collaboration],
  %``Search for $WZ$ resonances in the fully leptonic channel using $pp$ collisions at $\sqrt{s}$ = 8 TeV with the ATLAS detector,''
  Phys.\ Lett.\ B {\bf 737}, 223 (2014)
  [arXiv:1406.4456 [hep-ex]].
\bibitem{atlas2}
  G.~Aad {\it et al.} [ATLAS Collaboration],
  %``Search for resonant diboson production in the $\mathrm {\ell \ell }q\bar{q}$ final state in $pp$ collisions at $\sqrt{s} = 8$ TeV with the ATLAS detector,''
  Eur.\ Phys.\ J.\ C {\bf 75}, no. 2, 69 (2015)
  [arXiv:1409.6190 [hep-ex]].
\bibitem{atlas3}
  G.~Aad {\it et al.} [ATLAS Collaboration],
  %``Search for production of $WW/WZ$ resonances decaying to a lepton, neutrino and jets in $pp$ collisions at $\sqrt{s}=8$  TeV with the ATLAS detector,''
  Eur.\ Phys.\ J.\ C {\bf 75}, no. 5, 209 (2015)
  [Eur.\ Phys.\ J.\ C {\bf 75}, 370 (2015)]
  [arXiv:1503.04677 [hep-ex]].

\bibitem{Khachatryan:2014gha}
  V.~Khachatryan {\it et al.} [CMS Collaboration],
  %``Search for massive resonances decaying into pairs of boosted bosons in semi-leptonic final states at $\sqrt{s} =$ 8 TeV,''
  JHEP {\bf 1408}, 174 (2014)
  [arXiv:1405.3447 [hep-ex]].

\bibitem{Aad:2009wy}
  G.~Aad {\it et al.} [ATLAS Collaboration],
  %``Expected Performance of the ATLAS Experiment - Detector, Trigger and Physics,''
  arXiv:0901.0512 [hep-ex].

%\cite{Roe:2004na}
\bibitem{Roe:2004na}
  B.~P.~Roe, H.~J.~Yang, J.~Zhu, Y.~Liu, I.~Stancu and G.~McGregor,
  %``Boosted decision trees, an alternative to artificial neural networks,''
  Nucl.\ Instrum.\ Meth.\ A {\bf 543}, no. 2-3, 577 (2005)
  [physics/0408124].
  %%CITATION = PHYSICS/0408124;%%
  %148 citations counted in INSPIRE as of 09 sept. 2015

\end{thebibliography}
\end{document}